\documentclass{llncs}

\usepackage{listings}
\lstdefinestyle{customc}{
  belowcaptionskip=1\baselineskip,
  breaklines=true,
  frame=L,
  xleftmargin=\parindent,
  language=SQL,
  showstringspaces=false,
  basicstyle=\fontsize{7}{9}\ttfamily,
  keywordstyle=\bfseries\color{black!40!black},
  commentstyle=\itshape\color{purple!40!black},
  identifierstyle=\color{black},
  stringstyle=\color{orange},
}

\lstdefinestyle{customasm}{
  belowcaptionskip=1\baselineskip,
  frame=L,
  xleftmargin=\parindent,
  language=[x86masm]Assembler,
  basicstyle=\footnotesize\ttfamily,
  commentstyle=\itshape\color{purple!40!black},
}
\usepackage{textpos}
\usepackage{blindtext}
\usepackage{placeins}
\usepackage[dvipsnames]{xcolor}
\usepackage{amsthm}
\usepackage[francais,english]{babel}
\usepackage{framed}
\usepackage{booktabs}

\usepackage{url}
\usepackage[toc,page]{appendix}
\usepackage{multirow,array}

\usepackage{soul}

\usepackage[utf8]{inputenc}

\usepackage{amsfonts,amstext,amsmath,amssymb}

\usepackage{enumerate}

\usepackage{cancel}
\usepackage{caption}
\usepackage{graphicx}
\usepackage{subfig}
\usepackage{float}
\usepackage{graphics}
\usepackage{fancyhdr}
\usepackage{wrapfig}

\usepackage{amsmath}
\usepackage[linesnumbered,ruled,vlined]{algorithm2e}
\usepackage[noend]{algcompatible}

\newcommand{\ie}{\textit{i}.\textit{e}., }

\newcommand{\eg}{\textit{e}.\textit{g}., }
\newcommand{\Eg}{\textit{E}.\textit{g}., }
\newcommand{\etc}{\textit{etc}.}

\usepackage{listings}
\usepackage{tgcursor}
\usepackage{multirow}

\usepackage{color}
\usepackage{times}
\usepackage{soul}

\title{Strider: A Hybrid Adaptive Distributed RDF Stream Processing Engine}
\author{Xiangnan Ren$^{1,2}$, Olivier Curé$^{2}$}
\institute{	
 $^{1}$ ATOS - 80 Quai Voltaire, 95870 Bezons, France\\
 \email{xiang-nan.ren@atos.net}\\
 $^{2}$ LIGM (UMR 8049), CNRS, UPEM, F-77454, Marne-la-Vallée, France  \\
   \email{olivier.cure@u-pem.fr}\\
 }

\begin{document}
\maketitle

\vspace{-5mm}
\begin{abstract}
Real-time processing of data streams emanating from sensors is becoming a common task in Internet of Things scenarios. The key implementation goal consists in efficiently handling massive incoming data streams and  supporting advanced data analytics services like anomaly detection. In an on-going, industrial project, we found out that a $24/7$ available stream processing engine usually faces dynamically changing data and workload characteristics. These changes impact the engine's performance and reliability. We propose Strider, a hybrid adaptive distributed RDF Stream Processing engine that optimizes logical query plan according to the state of data streams. Strider has been designed to guarantee important industrial properties such as scalability, high availability, fault-tolerant, high throughput and acceptable latency. These guarantees are obtained by designing the engine's architecture with state-of-the-art Apache components such as Spark and Kafka. We highlight the efficiency (\eg on a single machine machine, up to 60x gain on throughput compared to state-of-the-art systems, a throughput of 3.1 million triples/second on a 9 machines cluster, a major breakthrough in this system's category) of Strider on real-world and synthetic data sets.

\end{abstract}

% \vspace*{-2mm}
\section{Introduction}
% \vspace*{-1mm}

%The vast amount of data produced by the Internet of Things (IoT)  generally needs to be processed in almost real-time. 
With the growing use of Semantic Web Technology in Internet of Things (IoT) contexts, \eg for data integration and reasoning purposes, the requirement for almost real-time platforms that can efficiently adapt to large scale data streams, \ie continuous SPARQL query processing, is gaining more and more attention. In the context of the FUI (Fonds Unique Interministeriel) Waves project\footnote{http://www.waves-rsp.org/}, we are processing data streams emanating from sensors distributed over the drinking water distribution network of a resource management international company. For France alone, this company distributes water to over 12 million clients through a network of more than 100.000 kilometers equipped with thousands (and growing) of sensors. Obviously, our RDF Stream Processing (RSP) engine should satisfy some common industrial features, \eg high throughput, high  availability, low latency, scalability and fault-tolerance.

Querying over RDF data streams could be quite challenging. Due to fast generation rates and schema free natures of RDF data streams, a continuous SPARQL query usually involves intensive join tasks which may rapidly become a performance bottleneck. Existing centralized RSP systems like C-SPARQL \cite{CSPARQL}, CQELS \cite{CQELS} and ETALIS \cite{ETALIS} are not capable of handling massive incoming data streams, as they do not benefit from task parallelism and the scalability of a computing cluster. Besides, most streaming systems are operating $24/7$ with patterns 
%(\eg number of temperature or flow observations)
,\ie stream graph structures, that may change overtime (in terms of graph shapes and sizes). This can potentially have a performance impact on query processing since in most available distributed RDF streaming systems, \eg CQELSCloud \cite{CQELSCloud} and Katts \cite{Katts}, the logical query plan is determined at compile time. Such a behavior can hardly promise long-term efficiency and reliability, since there is no single query plan that is always optimal for a given query. 
%Due to the high expertise entry point (distributed systems, database management systems and Semantic Web), we consider that an RSP system covers all previously mentioned aspects is still unavailable.

A general approach for large scale data stream processing is performed over a distributed settings. Such systems are better designed and operated upon when implemented on top of robust, state-of-the-art engines, \eg Kafka \cite{Kafka} and Spark \cite{Spark,Sparkstreaming}. Moreover, the system has to adapt to unpredictable input data streams and to dynamically updated execution plans while ensuring optimal performance. A time-driven/batch-driven \cite{SECRET} approach could be a solution for adaptive streaming query. In that context, it becomes possible to reconstruct the logical plan for each query execution. Furthermore, compared to data-driven systems \cite{SECRET}, time-driven/batch-driven provides a more coarse operation granularity. Although this mechanism inevitably causes higher query latency, it also brings high system throughput, inexpensive cost and low latency to achieve fault tolerance and system adaptivity \cite{Sparkstreaming}.

Our system, Strider, possesses the aforementioned characteristics. In this paper, we present three main contributions concerning this system: (1) the design and implementation of a production-ready RSP engine for large scale RDF data streams processing which is based on the state-of-the-art distributed computing frameworks (\ie Spark and Kafka). (2) Strider   integrates two forms of adaptation. In the first one, for each execution of a continuous query, the system decides, based on incoming stream volumes, to use either a query compile-time (rule-based) or query run-time (cost-based) optimization approach. The second one concerns the run-time approach and decides when the query plan is optimized (either at the previous query window or at the current one). (3) an evaluation of Strider over real-world and synthetic data sets.

%paper details our main contributions which concern the run-time approach. More precisely, our contributions are:}
%In summary, our major contributions consist of: (1) the design and implementation of a production-ready RSP engine for large scale RDF data streams processing which is based on the state-of-the-art distributed computing frameworks (\ie Spark and Kafka). (2) The integration of an adaptive stream-based query optimizer with a set of optimization strategies. This is achieved by composing between two optimization techniques, \ie heuristic and cost-based, and two optimization processing timing, \ie at the current time window (denoted \emph{forward}) and during the previous window (\emph{backward}). The system dynamically decides which of the four solutions is the most adapted to a certain situation.
%We achieve the adaptivity of the query optimizer by proposing two rule-based, cost-based and so-called \emph{backward/forward} strategies. 
%(3) An evaluation of Strider over real-world and synthetic data sets.

% \vspace{-2mm}
\section{Background Knowledge}
% \vspace{-1mm}

Strider follows a classical streaming system approach with a messaging component for data flow management and a computing core for real-time data analytics. In this section, we present and motivate the use of Spark Streaming and Kafka as these two components. Then, we consider streaming models and adaptive query processing.

\textbf{Kafka \& Spark Streaming.} 
Kafka is a distributed message queue which aims to provide a unified, high-throughput, low-latency real-time data management. Intuitively, producers emit messages which are categorized into adequate \emph{topics}. The messages are partitioned among a cluster to support parallelism of upstream/downstream operations. Kafka uses \emph{offsets} to uniquely identify the location of each message within the partition. 
%The offset is managed by Kafka itself or Zookeeper to ensure zero information loss and high consistency of data.

Spark is a MapReduce-like cluster-computing framework that proposes a parallelized fault-tolerant collection of elements called Resilient Distributed Dataset (RDD) \cite{Spark}. An RDD is divided into multiple partitions across different cluster nodes such that operations can be performed in parallel. Spark enables parallel computations on unreliable machines and automatically handles locality-aware scheduling, fault-tolerant and load balancing tasks. 
%The computation is described as a Directed Acyclic Graph (DAG) of operators and is partitioned into different \emph{stages}. 
Spark Streaming extends RDD to Discretized Stream (DStream) \cite{Sparkstreaming} and thus enables to support near real-time data processing by creating \emph{micro-batches} of duration $T$. DStream represents a sequence of RDDs where each RDD is assigned a timestamp. Similar to Spark, Spark Streaming describes the computing logics as a template of RDD Directed Acyclic Graph (DAG). Each batch generates an instance according to this template for later job execution. The micro-batch execution model provides Spark Streaming second/sub-second latency and high throughput. To achieve continuous SPARQL query processing on Spark Streaming, we bind the SPARQL operators to the corresponding Spark SQL relational operators. Moreover, the data processing is based on DataFrame (DF), an API abstraction derived from RDD. 

\textbf{Streaming Models.} At the physical level, a computation model for stream processing has two principle classes: Bulk Synchronous Parallel (BSP) and Record-at-a-time \cite{Drizzle}. From a logical level perspective, a streaming model uses the concept of a \emph{Tick} to drive the system in taking actions over input streams. \cite{SECRET} defines a Tick in three ways: data-driven (DD), time-driven (TD) and batch-driven (BD). In general, the physical BSP is associated to the TD and/or BD models, \eg Spark Streaming\cite{Sparkstreaming} and Google DataFlow with FlumeJava \cite{DataFlow} adopt this approach by creating a micro-batch of a certain duration $T$. That is data are cumulated and processed through the entire DAG within each batch.  The record-at-a-time model is usually associated to the logical DD model (although TD and BD are possible) and  prominent examples are Flink \cite{Flink} and Storm \cite{Storm}.
% \textbf{Streaming Models.} \textcolor{blue}{At the physical level, a computation model for stream processing has two principle classes: Bulk Synchronous Parallel (BSP) and Record-at-a-time \cite{Drizzle}. From the logical level perspective, a streaming model uses the concept of a \emph{Tick} to drive the system to take action over input streams \cite{SECRET}. This paper defines a Tick in three ways: data-driven, time-driven and batch-driven. Systems such as Spark Streaming\cite{Sparkstreaming} and Google DataFlow with FlumeJava \cite{DataFlow} adopt BSP with both the time and batch-driven approaches and perform stream processing by creating a micro-batch of a certain duration $T$. That is data are cumulated and processed through the entire DAG within each batch. An alternative stream processing system implementation approach is presented in Flink \cite{Flink} and Storm \cite{Storm} and uses a so called record-at-a-time mechanism, where operators maintain mutable states. }
%As a result, we propose a simple comparison for both physical and logical streaming models. 
The record-at-a-time/DD model provides lower latency than BSP/TD/BD model for typical computation. On the other hand, the record-at-a-time model requires state maintenance for all operators with record-level granularity. This behavior obstructs system throughput and brings much higher latencies when recovering after a system failure \cite{Drizzle}. For complex tasks involving lots of aggregations and iterations, the record-at-a-time model could be less efficient, since it introduces an overhead for the launch of frequent tasks. Given these properties and the fact that in \cite{Facebook}, the authors  emphasize that latencies in the order of few seconds is enough for most extreme use cases at Facebook, we have decided to use Spark Streaming.

\textbf{Adaptive Query Processing (AQP)} is recognized as a complex task, especially in the streaming context \cite{AQP}. Moreover, AQP for continuous SPARQL query needs to cope with some cross-field challenges such as SPARQL query optimization, stream processing, \etc. Due to structure unpredictability, schema-free and real-time features of RDF data streams, conventional optimizations for static RDF data processing through data pre-processing, \eg triple indexing and statistic summarizing, become impractical. However, the perspectives from \cite{Emergent_Schemas,sparql2sql} show that most parts of RDF graphs have tabular structure, especially in the IoT domain. 
This opens up several perspectives concerning  selectivity/cardinality estimation and the possibility to use Dynamic Programming (DP) approaches. Inspired by \cite{Jena,Thomas-CS,Thomas-EDBT,HP,Drizzle}, we propose a novel AQP optimizer for RDF stream processing.

\vspace{-3mm}
\section{Strider overview}
\label{sec:Strider}
\vspace{-2mm}
In this section, we first present a Strider query example, then we provide a system's overview, detail the data flow and query optimization components.

\vspace{-2mm}
\subsection{Continuous query example} 
\vspace{-2mm}
Listing \ref{Q1} introduces a running scenario that we will use throughout this paper. The example corresponds to a use case encountered in the Waves project, \ie query $Q_8$ continuously processes the messages of various types of sensor observations. 

We introduce new lexical rules for continuous SPARQL queries which are tailored to a micro-batch approach.The  \texttt{STREAMING} keyword initializes the application context of Spark Streaming and the windowing operator. More precisely, \texttt{WINDOW} and \texttt{SLIDE} respectively indicate the size and sliding parameter of a time-based window. The novelty comes from the \texttt{BATCH} clause which specifies the micro-batch interval of discretized stream for Spark Streaming. Here, a sliding window consists of one or multiple micro-batches.

\vspace{1mm}
\begin{lstlisting}[frame=single,caption=Strider's query example ($Q_8$), label=Q1,captionpos=b]
STREAMING { WINDOW [10 Seconds] SLIDE [10 Seconds] BATCH [5 Seconds] }
REGISTER { QUERYID [Q8] SPARQL [ 
  prefix rdf: <http://www.w3.org/1999/02/22-rdf-syntax-ns#>
  prefix ssn: <http://purl.oclc.org/NET/ssnx/ssn/>
  prefix cuahsi: <http://www.cuahsi.org/waterML/>
  SELECT ?s ?o1 ?o2 ?o3
  WHERE {   ?s ssn:hasValue ?o1 (tp1); ssn:hasValue ?o2 (tp2); 
               ssn:hasValue ?o3 (tp3).
            ?o1 rdf:type cuahsi:flow (tp4). 
            ?o2 rdf:type cuahsi:temperature (tp5).
            ?o3 rdf:type cuahsi:chlorine    (tp6). }] }
\end{lstlisting}
\vspace{-2.5mm}

The \texttt{REGISTER} clause is used to register standard SPARQL queries. Each query is identified by an identifier. The system allows to register several queries simultaneously in a thread pool. By sharing the same application context and cluster resources, Strider launches all registered continuous SPARQL queries asynchronously by different threads.

\vspace{-2mm}
\subsection{Architecture}
\label{Architecture_Overview}
\vspace{-1mm}

Strider contains two principle modules: (1) data  flow management. In order to ensure high throughput, fault-tolerance, and easy-to-use features, Strider uses Apache Kafka to manage input data flow. The incoming RDF streams are categorized into different \emph{message topics,} which practically represent different types of RDF events. (2) Computing core. Strider core is based on the Spark programming framework. Spark Streaming receives, maintains messages emitted from Kafka in parallel, and generates data processing pipeline.

\begin{figure}[h]
\vspace{-5mm}
\begin{center}
\includegraphics[width=1\linewidth]{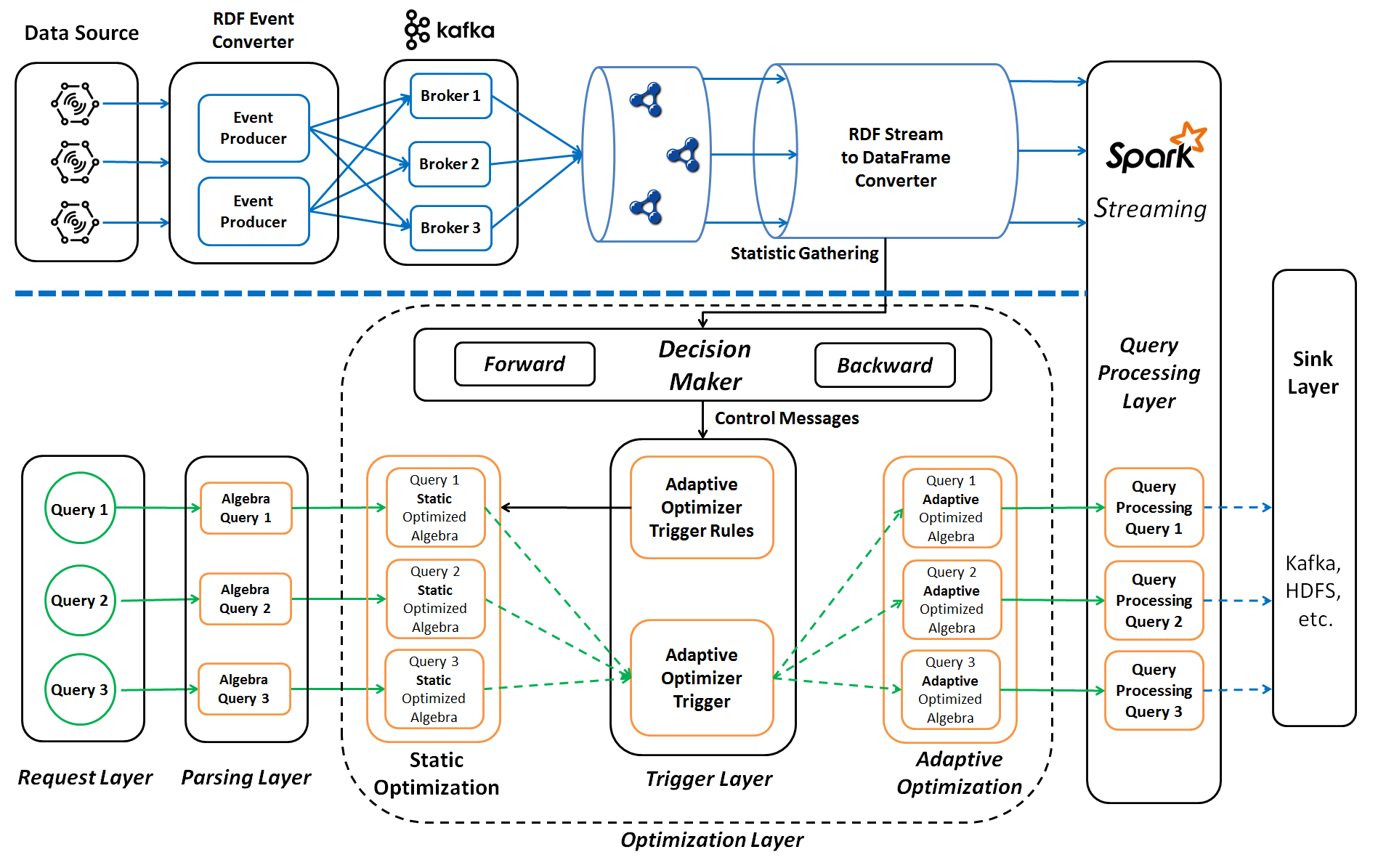}
\captionof{figure}{Strider Architecture}
\label{Strider_Architecture}
\end{center}
\vspace{-7mm}
\end{figure}

Figure \ref{Strider_Architecture} gives a high-level overview of the system's architecture. The upper part of the figure provides  details on the application's data flow management. In a nutshell, data sources (IoT sensors) are sending messages to a publish-subscribe layer. This layer emits messages for the streaming layer which executes registered queries. The sensor's metadata are converted into RDF events for data integration purposes. We use Kafka to design the system's data flow management. Kafka is connected to Spark Streaming using a \emph{Direct Approach}\footnote{https://spark.apache.org/docs/latest/streaming-kafka-integration.html} to guarantee exactly-once semantics and parallel data feeding. The input RDF event streams are then continuously transformed to DataFrames.

The lower part of Figure \ref{Strider_Architecture} presents  components related to the implementation of the computing core. The Request layer registers continuous queries. Currently, we consider that the input queries are independent, thus a multi-query optimization approach (\eg sub-query sharing) is not in the scope of the current state of Strider. These queries are later sent to the Parsing layer to compute a first version of a query plan. These new plans are pushed to the Optimization layer which consists of four collaborating sub-components: static and adaptive optimizations as well as a trigger mechanism and a Decision Maker for adaptation strategy. Finally, the Query Processing layer sets the query execution off right after the optimized logical plan takes place. 

\vspace{-2mm}
\section{Strider's continuous SPARQL processing}
\vspace{-1mm}

In this section, we detail the components of the Strider's optimizer layer. Two optimization components are proposed, \ie static and adaptive, which are respectively based on heuristic rules and (stream-based) statistics.
The trigger layer decides whether the query processing adopts a static or an adaptive approach.
Two strategies are proposed for AQP: backward (B-AQP) and forward (F-AQP). They mainly differ on  when, \ie at the previous or current window, the query plan is computed.

\vspace{-2mm}
\subsection{Query processing outline \& trigger layer} 
\vspace{-1mm}
Intuitively, Strider's optimizers search for the optimal join ordering of triple patterns based on collected statistics.
%\st{Strider's optimizers aim at reconstructing a SPARQL query's  Basic Graph Pattern (BGP) at runtime. That is the system searches for the optimal join ordering of triple patterns based on collected statistics.} 
Both static (query compile-time) and adaptive (query run-time) optimizations are processed using a graph $G^U = (V,E)$, denoted Undirected Connected Graph (UCG) \cite{Jena} where vertices represent triple patterns and edges symbolize joins between triple patterns. Naturally, for a given query $q$ and its query graph $G^Q(q)$, $G^U(q) \subseteq G^Q(q)$. A UCG showcases the structure of a BGP and the join possibilities among its triple patterns. That query representation is considered to be more expressive \cite{Thomas-CS} than the classical RDF query graph. The weight of UCG's vertices and edges correspond to the selectivity of triple patterns and joins, respectively. Once the weights of an UCG are initialized, the query planner automatically generates an optimal logical plan and triggers a query execution. For the sake of a better explanation, the windowing operator involved in this section is considered as a tumbling window.

Strider's static optimization retains the philosophy of \cite{HP}. Basically, static optimization implies a heuristics-based query optimization. It ignores data statistics and leads to a static query planner. In this case, unpredictable changes in data  stream structures may incur a bad query plan. The static optimization layer aims at giving a basic performance guarantee. 
%In this layer, the optimizer first creates a UCG graph using a set of heuristic rules. 
The predefined heuristic rules set empirically assign the weights for UCG vertices and edges. Next, the query planner determines the shortest traversal path in the current UCG and generates the logical plan for query execution. The obtained logical plan represents the query execution pipeline which is permanently kept by the system. More details about UCG creation and query logical plan generation are given in Sec. \ref{subsec:Insight_Into_AQP}.

The Trigger layer supports the transition between the stages of static optimization and adaptive optimization. In a nutshell, that layer is dedicated to notify the system whether it is necessary to proceed an  adaptive optimization. Our adaptation strategy requires collecting statistical information and generating an execution logical plan. The overhead coming with such actions is not negligible in a distributed environment. The Strider prototype provides a set of straightforward trigger rules, \ie the adaptive algebra optimization is triggered by a configurable workload threshold. The threshold refers to two factors: (1) the input number of RDF events/triples; (2) the fraction of the estimated input data size and the allocated executors' heap memory.

\vspace{-2mm}
\subsection{Run-time query plan generation}
\label{subsec:Insight_Into_AQP}
\vspace{-1mm}

Here, we first briefly introduce how we collect stream statistics and construct query plan. Then, we give an insight into the AQP optimization, which is essentially a cardinality-based optimization. 

\begin{figure}[h]
\vspace{-4.5mm}
\advance\leftskip-0.55cm\includegraphics[width=1.07\linewidth]{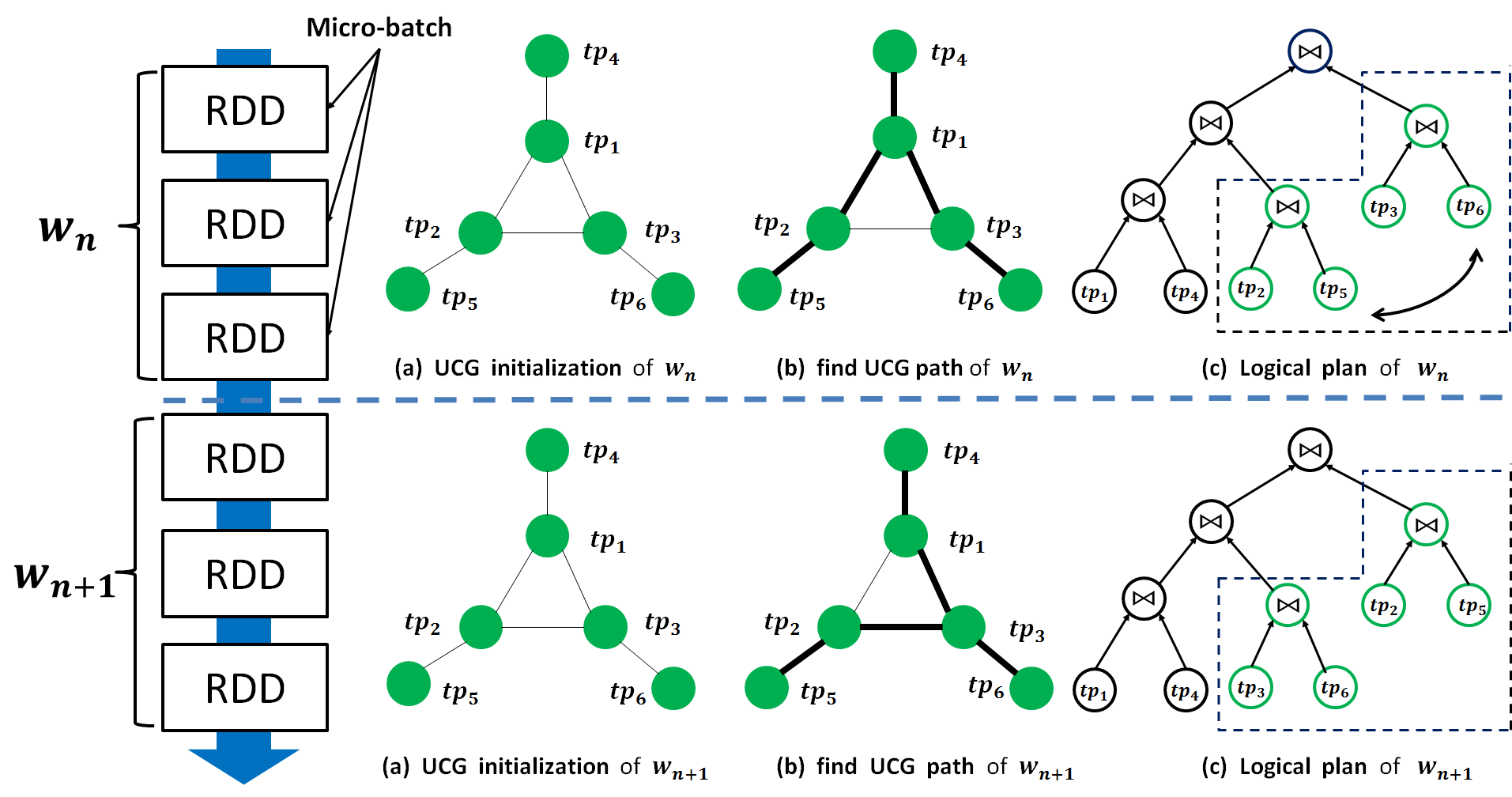}
\captionof{figure}{Dynamic Query Plan Generation for $Q_8$}
\label{fig:Dynamic_Optim}
\vspace{-4.5mm}
\end{figure}

Unlike systems based on greedy and left-deep tree generation, \eg \cite{Jena,CQELS}, Strider makes a full usage of CPU computing resources and benefits from parallel hardware settings. It thus creates query logical plans in the form of general (bushy) directed trees. Hence, the nodes with the same height in a query plan $p_n$ can be asynchronously computed in a non-blocking way (in the case where computing resources are allowed). Coming back to our Listing \ref{Q1} example, Figure \ref{fig:Dynamic_Optim} refines the procedure of query processing (F-AQP) at $w_{n}$, $n \in N$. If $w_n$ contains multiple RDDs (micro-batches), the system performs the union all RDDs and generates a new combined RDD. Note that the union operator has a very low-cost in Spark. Afterward, the impending query plan optimization follows three steps: (a) UCG (weight) initialization; (b) UCG path cover finding; (c) query plan generation.  

\begin{figure}[h]
\vspace{-2mm}
\advance\leftskip-0.25cm\includegraphics[width=1\linewidth]{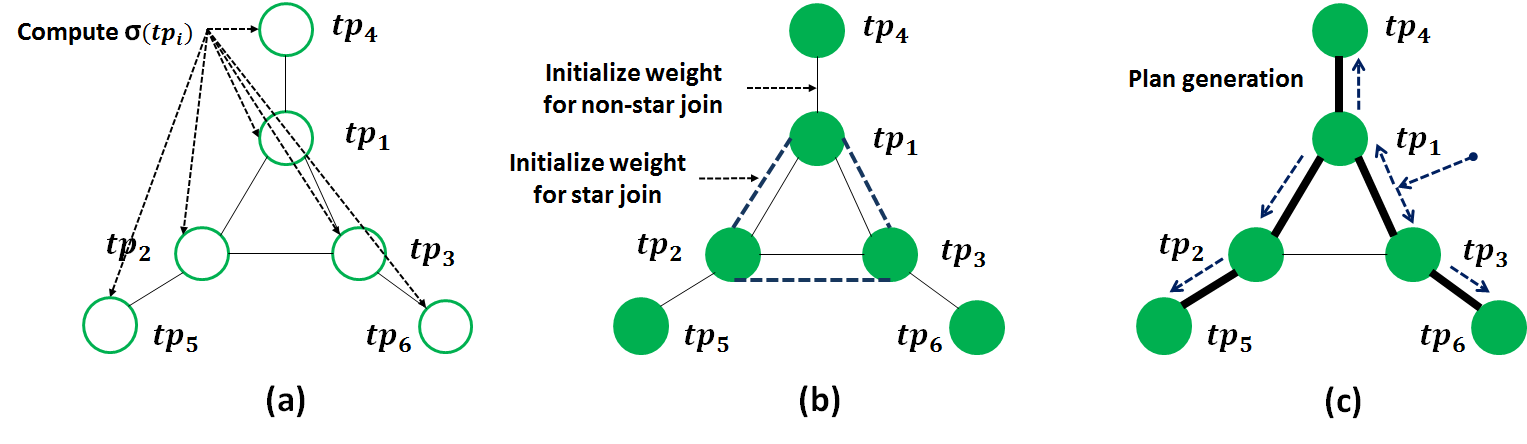}
\captionof{figure}{Initialized UCG weight, find path cover and generate query plan}
\label{fig:path_finding}
\vspace{-6mm}
\end{figure}

\textbf{ UCG weight initialization} is briefly described in Algorithm \ref{algo:ucg_init} and Figure \ref{fig:path_finding} (step (a), step (b)).
%\textcolor{red}{Maybe to delete: For a given graph $G^Q(q)$ of query $q$, we have $\forall G^U(q) \subseteq G^Q(q)$}. 
Since triple patterns are located at the bottom of a query tree, the query evaluation is performed in a bottom-up fashion and starts with the selection of triple patterns $\sigma(tp_i)$, $1 \leq i \leq I$ (with $I$ the number of triple patterns in the query's BGP). The system computes $\sigma(tp_i)$ asynchronously for each $i$ and temporally caches the corresponding results ($R^{\sigma}(tp_i)$) in memory. $Card(tp_i)$, \ie the cardinality of $R^{\sigma}(tp_i)$, is computed by a Spark count action. Thence, we can directly assign the weight of vertices in $G^U(Q)$. Note that the estimation of $Card(tp_i)$ is exact. 

Once all vertices are set up, the system predicts the weight of edges (\ie joined patterns) in $G^U(q)$. We categorize two types of joins (edges): (i) star join, includes two sub-types, \ie star join without bounded object and star join with bounded object; (ii) non-star join. 
To estimate the cardinality of join patterns, we make a trade-off between accuracy and complexity. The main idea is inspired by a research conducted in \cite{Jena,Thomas-CS,Thomas-EDBT}. However, we infer the weight of an edge from its connected vertices, \ie no data pre-processing is required. The algorithm begins by iteratively traversing $G^U(q)$ and identifies each vertex $v \in V$ and each edge $e \in E$. Then we can decompose $G^U(q)$ into the disjoint star-shaped joins and their interconnected chains (Figure \ref{fig:path_finding}, step (b)). The weight of an edge in a star join shape is estimated by the function getStarJoinWeight. The function first estimates the upper bound of each star join output cardinality (\eg, $Card(tp_1 \bowtie tp_2 \bowtie tp_3)$), then assigns the weight edge by edge. Every time the weight of the current edge $e$ is assigned, we mark $e$ as visited. This process repeats until no more star join can be found. Then, the weight of unvisited non-star join shapes is estimated by the function getNonStarJoinweight. It lookups the two vertices of the current edge, and chooses the one with smaller weight to estimate the edge cardinality. The previous processes are repeated until all the edges have been visited in $G^U(q)$. 

% \begin{figure}[h]
% \vspace{-2mm}
% \advance\leftskip-0.25cm\includegraphics[width=1\linewidth]{WeightInit.png}
% \captionof{figure}{Initialized UCG weight, find path cover and generate query plan}
% \label{fig:path_finding}
% \vspace{-6mm}
% \end{figure}

% \vspace{-6mm}
% \begin{algorithm}[ht!]
% \caption{UCG weight initialization}
% \label{algo:ucg_init}
% \KwIn{query $q$, $G^U(q) = (V,E) \subseteq G^Q(q)$, current buffered window $w_n$}
% \KwOut{$G^U(q)$ with weight-assigned}

% \SetKwFunction{compute}{compute}
% \SetKwFunction{buffer}{buffer}
% \SetKwFunction{getStarJoinWeight}{getStarJoinWeight}
% \SetKwFunction{getNonStarJoinWeight}{getNonStarJoinWeight}

% \While{$\exists v$ unvisited $\in V$ }{
% mark $v$ as visited, 
% % compute the vertex $v$, 
% $R^{\sigma}(v) \gets $ \compute($v$) \;
% \buffer ($v$, $R^{\sigma}(v)$) $\wedge$ $v$.weight $\gets Card(v)$ \;
% }

% \While{$\exists e$ unvisited $\in E$}{
% mark $e$ as visited \;
% \If{($\exists$ star join $S_J$) $\wedge e \cap S_J  \neq \emptyset$}{
% locate each $S_J \in G^U(q)$ \\
% \ForEach{$\forall e_S \in S_J$}{
% mark $e_S$ as visited \;
% $e_S$.weight $\gets \getStarJoinWeight(S_J, e_S.vertices)$ \;}
% }
% \lElse{ $e$.weight $\gets \getNonStarJoinWeight(S_J)$}
% }
% \end{algorithm}
% \vspace{-5.5mm}

\textbf{UCG path cover finding \& Query plan generation.} Figure \ref{fig:path_finding} step (c) introduces path cover finding and query plan generation. The system starts by finding the path cover in $G^U(q)$ right after $G^U(q)$ is prepared. Intuitively, we search the undirected path cover which links all the vertices of $G^U(q)$ with a minimum total edge weight. The path searching is achieved by applying Floyd–Warshall algorithm iteratively. The extracted path $Card(G^U(q)) \subseteq G^U(q)$, is regarded as the candidate for the logical plan generation. Finally, we construct $p_n$, the logical plan of $G^U(q)$ at $w_n$, in a top-down manner (Figure \ref{fig:path_finding}, step (c)). Note that path finding and plan generation are both computed  on the driver node and are not expensive operations (around 2 - 4 milliseconds in our case).

\vspace{-6mm}
\begin{algorithm}[ht!]
\caption{UCG weight initialization}
\label{algo:ucg_init}
\KwIn{query $q$, $G^U(q) = (V,E) \subseteq G^Q(q)$, current buffered window $w_n$}
\KwOut{$G^U(q)$ with weight-assigned}

\SetKwFunction{compute}{compute}
\SetKwFunction{buffer}{buffer}
\SetKwFunction{getStarJoinWeight}{getStarJoinWeight}
\SetKwFunction{getNonStarJoinWeight}{getNonStarJoinWeight}

\While{$\exists v$ unvisited $\in V$ }{
mark $v$ as visited, 
% compute the vertex $v$, 
$R^{\sigma}(v) \gets $ \compute($v$) \;
\buffer ($v$, $R^{\sigma}(v)$) $\wedge$ $v$.weight $\gets Card(v)$ \;
}

\While{$\exists e$ unvisited $\in E$}{
mark $e$ as visited \;
\If{($\exists$ star join $S_J$) $\wedge e \cap S_J  \neq \emptyset$}{
locate each $S_J \in G^U(q)$ \\
\ForEach{$\forall e_S \in S_J$}{
mark $e_S$ as visited \;
$e_S$.weight $\gets \getStarJoinWeight(S_J, e_S.vertices)$ \;}
}
\lElse{ $e$.weight $\gets \getNonStarJoinWeight(S_J)$}
}
\end{algorithm}
\vspace{-5.5mm}

\vspace{-2mm}
\subsection{B-AQP \& F-AQP}
\label{BFAQP}
\vspace{-1mm}

\begin{figure}[h]
\vspace{-5.5mm}
\begin{center}
\advance\leftskip-0.2cm
\includegraphics[width=1.05\linewidth]{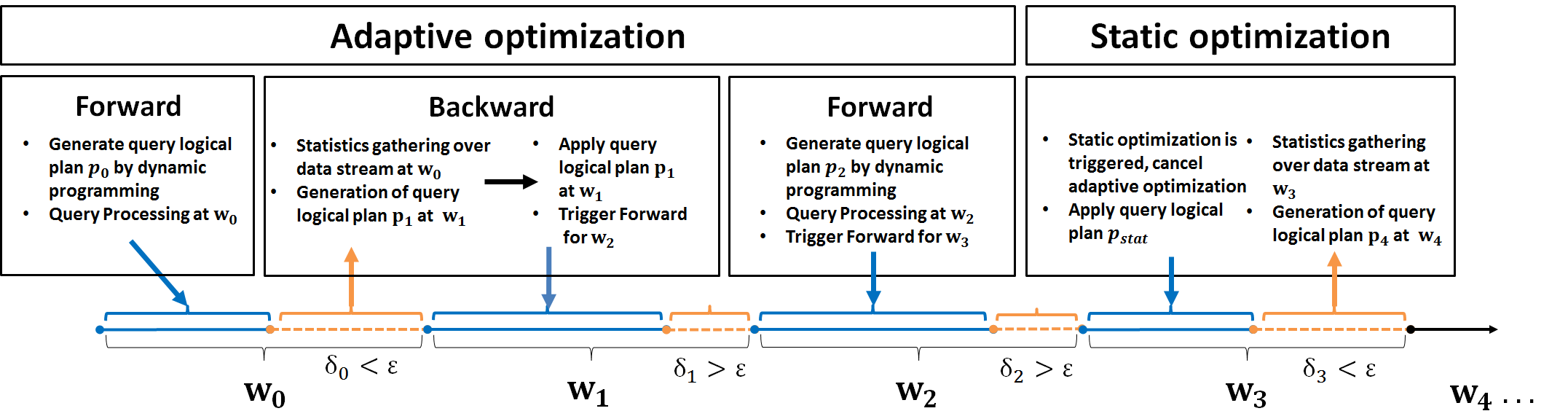}
\captionof{figure}{Decision Maker of Adaptation Strategy}
\label{Decision_Maker}
\end{center}
\vspace{-9.5mm}
\end{figure}

We propose a dual AQP strategy, namely, \textbf{backward} (B-AQP) and \textbf{forward} (F-AQP). B/F-AQP depict two philosophies for AQP, Figure \ref{Decision_Maker} roughly illustrates how B/F-AQP switching is decided at run-time, \ie this is the responsibility of the Decision Maker component. Generally, B-AQP and F-AQP are using similar techniques for query plan generation. Compared to F-AQP, B-AQP delays the process for query plan generation.

% \begin{figure}[h]
% \vspace{-4.5mm}
% \begin{center}
% \advance\leftskip-0.2cm
% \includegraphics[width=1.05\linewidth]{AQP_chain.png}
% \captionof{figure}{Decision Maker of Adaptation Strategy}
% \label{Decision_Maker}
% \end{center}
% \vspace{-9.5mm}
% \end{figure}

Our B-AQP strategy is inspired by \cite{Drizzle}'s pre-scheduling. Backward implies gathering, feeding back the statistics to the optimizer on the current window, then the optimizer constructs the query plan for the next window. That is the system computes the query plan $p_{n+1}$ of a window $w_{n+1}$ through the statistics of a previous window $w_n$. Strider possesses a time-driven execution mechanism, the query execution is triggered periodically with a fixed update frequency $s$ (\ie sliding window size). Between two consecutive window $w_n$ and $w_{n+1}$, there is a computing barrier to reconstruct the query plan for $w_{n+1}$ based on the collected statistics from a previous window $w_n$. Suppose the query execution of $w_n$ consumes a time $t_n$ (\eg in seconds), then for all $t_n < s$, the idle duration $\delta_n = s-t_n$ allows to re-optimize the query plan. But $\delta_n$ should be larger than a configurable threshold $\epsilon$. For $\delta_n < \epsilon$, the system may not have enough time to (i) collect the statistic information of $w_n$ and (ii) to construct a query plan for $w_{n+1}$. This potentially expresses a change of incoming steams and a degradation of query execution performance. Hence, the system decides to switch to the F-AQP approach.

F-AQP applies a DP strategy to find the optimal logical query  plan for the current window $w_n$. The main purpose of F-AQP is to adjust the system state as soon as possible. The engine executes a query, collects statistics and computes the logical query plan simultaneously. Here, the statistics are obtained by counting intermediate query results, which causes data shuffling and DAG interruption, \ie the system has to temporally cut the query execution pipeline. In Spark, such suspending operation is called an \emph{action}, which immediately triggers a job submission in Spark application. However, frequent job submission may bring some side effects. The rationale is, for a master-slave based distributed computing framework (\eg Spark, Storm) uses a master node (\ie driver) to schedule jobs.  The driver locally computes and optimizes each submitted DAG and returns the control messages to each worker node for parallel processing. Although the ``count" action itself is not expensive, the induced side effects (\eg driver job-scheduling/submission, communication of control message between driver and workers) will potentially impact the system's stability. For instance, based on our experience, F-AQP's frequent job submission and intermediate data persistence/unpersistence put a great pressure on the JVM's Garbage Collector (GC), \eg untypical GC pauses are observed from time to time in our experiment.

\textbf{Decision Maker.} Through experimentations of different Strider configurations, we understood the complementarity of both the B-AQP and F-AQP approaches. Real performance gains can be obtained by switching from one approach to another. This is mainly due to their properties which are summarized in Table \ref{tab:table1}.

\begin{table}[h!]
  \vspace{-5mm}
\centering
% \caption{B/F-AQP summarization}
  \begin{tabular}{ccc}
    \toprule
    \textbf{Strategy} & \textbf{Advantage} & \textbf{Drawback}\\
    \midrule
    \textbf{B-AQP} & { }{ }{ }{ } No dynamic programming overhead { }{ }{ }{ }
                   &{ }{ }{ }{ } Approximate query plan generation { }{ }{ }{ } \\
                   &  & { }{ }{ }{ } through previously-collected statistics { }{ }{ }{ } \\
    \midrule
    \textbf{F-AQP} & { }{ }{ }{ } Query plan generation through { }{ }{ }{ }
                   &{ }{ }{ }{ } Overhead for dynamic programming, { }{ }{ }{ } \\
                   & real-time collected statistics  &  side-effects caused by pipeline interruption\\
    \bottomrule
  \end{tabular}
      \caption{B/F-AQP summarization}
        \label{tab:table1}
\vspace{-9mm}
\end{table}

We designed a decision maker to automatically select the most adapted strategy for each query execution. The decision maker takes into account two parameters: a configurable switching threshold $\epsilon \in~]0,1[$; $\gamma_n = \frac{t_n}{s}$, the fraction of query execution time $t$ over windowing update frequency $s$. For the query execution at $w_n$, if $\gamma_n < \epsilon$, the system updates the query plan from $p_n$ to $p_{n+1}$ for the next execution. Otherwise, the system recomputes $p_{n+1}$ by DP at $w_{n+1}$ (see Algorithm \ref{algo:b_f_aqp}). We empirically set $\epsilon = 0.7$ by default. 

\vspace*{-6mm}
\begin{algorithm}[ht!]
\caption{B-AQP and F-AQP Switching in Decision Maker}
\label{algo:b_f_aqp}
\KwIn{query $q$, switching threshold $\epsilon$, sliding window $W = \{w_n\}_{n \in N}$,\quad \quad \quad  \\ \quad \quad \quad update frequency $s$ of $W$ }

\SetKwFunction{getRuntime}{getRuntime}
\SetKwFunction{getAdaptiveStrategy}{getAdaptiveStrategy}
\SetKwFunction{update}{update}
\SetKwFunction{execute}{execute}

\ForEach{$w_n \in W$ }{
$t_n \gets $ \getRuntime \{ \execute($q$) \} // executionTime \;
$\lambda_n \gets $ \getAdaptiveStrategy($\epsilon$,$t_n$,$s$) // adaptiveStrategy\;
\If{$\lambda_n$ == Backward}{ 
    update query plan $p_n$ of $q$ at $w_n$ \\
    $p_{n+1} \gets$  \update($p_n$)\;
         }
\lIf{$\lambda_n$ == Forward}{Recompute $p_{n+1}$ at $w_{n+1}$} 
}
\end{algorithm}
\vspace*{-6mm}

The decision maker plays a key role for maintaining the stability of the system's performance. Our experiment (Sec. \ref{subsec:eval}) shows that, the combination of F/B-AQP through decision maker is able to prevent the sudden performance declining during a long running time. 

\vspace*{-2mm}
\section{Evaluation}
\vspace*{-2mm}

\subsection{Implementation details}

Strider is written in Scala, the code source can be found here\footnote{https://github.com/renxiangnan/strider}.
To enable SPARQL query processing on Spark, Strider parses a query with Jena ARQ and obtains a query algebra tree in the Parsing layer. The system  reconstructs the algebra tree into a new Abstract Syntax Tree (AST) based on the Visitor model. Basically, the AST represents the logical plan of a query execution. Once the AST is created, it is pushed into the algebra Optimization layer. By traversing the AST, we bind the SPARQL operators to the corresponding Spark SQL relational operators for query evaluation.
% We use a series of micro-benchmarks to measure the performance of Strider, including the system's adaptivity. We first focus on continuous SPARQL query processing with stable stream structure. \Ie, in the ideal case, incoming data streams maintain invariant structure, the proportion of variant types of RDF triples does not change over time. We next demonstrate the efficiency of Strider's AQP by feeding the system structurally unstable RDF streams, \ie the structure of input stream varies over time. To that end, we first present the experimental setup and then provide results.

\vspace{-2mm}
\subsection{Experimental Setup}
\label{subsec:exp_setup}
\vspace{-1mm}

We test and deploy our engine on Amazon EC2/EMR cluster of 9 computing nodes and Yarn resource management. The system holds 3 nodes of m4.xlarge for data flow management (\ie Kafka broker and Zookeeper \cite{Zookeeper}). Each node has 4 CPU virtual cores of 2.4 GHz Intel Xeon E5-2676, 16 GB RAM and 750 MB/s bandwidth. We use Apache Spark 2.0.2, Scala 2.11.7 and Java 8 as baselines for our evaluation. The Spark (Streaming) cluster is configured with 6 nodes (1 master, 5 workers) of type c4.xlarge. Each one has 4 CPU virtual cores of 2.9 GHz Intel Xeon E5-2666, 7.5 GB RAM and 750 MB/s. The experiments of Strider on local mode, C-SPARQL and CQELS are all performed on a single instance of type c4.xlarge.

\textbf{Datasets \& Queries.} We evaluated our system using two datasets that are built around real world streaming use cases: \emph{SRBench} \cite{SRBench} and \emph{Waves}. SRBench, one of the first available RSP benchmarks, comes with 17 queries on LinkedSensorData. The datasets consists of weather observations about hurricanes and blizzards in the United States (from 2001 to 2009). Another dataset considered in our evaluation comes from aforementioned industrial project Waves. The dataset describes different water measurements captured by sensors. Values of flow, water pressure and chlorine levels are examples of these measurements. The value annotation uses three popular ontologies: SSN, CUAHSI-HIS and QUDT. Each sensor observes and records at least one physical phenomenon or a chemical property, and thus generates RDF data stream through Kafka producer. Our micro-benchmark contains 9 queries, denoted from $Q_1$ to $Q_9$\footnote{Check the wiki of our github page for more details of the queries and datasets}. The road map of our evaluation is designed as follow: (1) injection of structurally stable stream for experiment of $Q_1$ to $Q_6$. $Q_1$ to $Q_3$ are tested by SRBench datasets. Here, a comparison between Strider and the state of the art RSP systems \eg C-SPARQL and CQELS are also provided. Then we perform $Q_4$ to $Q_6$ based on Waves dataset. (2) Injection of structurally unstable stream. We generate RDF streams by varying the proportion of different types of Kafka messages (\ie sensor observations). For this part of the evaluation, queries $Q_7$ to $Q_9$ are considered. 

\textbf{Performance criteria.} In accordance with \emph{Benchmarking Streaming Computation Engines at Yahoo!} \footnote{https://yahooeng.tumblr.com/post/135321837876/benchmarking-streaming-computation-engines-at}, we choose the system throughput and query latency as two primary performance metrics. Throughput indicates how many data can be processed in a unit of time. Throughput is denoted as ``triples per second" in our case. Latency means how long does the RSP engine consumes between the arrival of an input and the generation of its output. The reason why we abandoned existing RSP performance benchmarking systems \cite{LSBench,City} is that, none of them is tailored for massive data stream. This limitation  is contrary to our original intention of using distributed stream processing framework to cope with massive RDF stream. We did not record the latency of C-SPARQL, CQELS and Strider in local mode for two reasons: (1) given the scalability limitation of C-SPARQL, we have to control input stream rate within a low level to ensure the engine can run normally \cite{LSBench}. (2) due to its design, based on a so-called eager execution mechanism and \emph{DStream} R2S operator, the measure of latencies in CQELS is unfeasible \cite{LSBench}. Moreover, given reasons provided in Sec. \ref{BFAQP}, we have not done any comparisons of B/F-AQP versus F-AQP approaches.

\textbf{Performance tuning} on Spark is quite difficult. Inappropriate cluster configuration may seriously hinder engine performance. So far we can only empirically configure Spark cluster and tune the cluster settings step by step. We briefly list some important performance settings based on our experience. First of all, we apply some basic optimization techniques. \eg using Kryo serializer to reduce the time for task/data serialization. Besides, we generally considered adjustments of Spark configuration along three control factors to achieve better performance. The first factor is the size of micro-batch intervals. Smaller batch sizes can better meet real-time requirements. However, it also brings frequent job submissions and job scheduling. The performance of a BSP system like Spark is sensitive to the chosen size of batch intervals. The second factor is GC tuning. Set appropriately, the GC strategy (\eg using Concurrent Mark-Sweep) and storage/shuffle fraction may efficiently reduce GC pressure. The third factor is the parallelism level. This includes the partition number of Kafka messages, the partition number of RDD for shuffling, and the upper/lower bound for concurrent job submissions, \etc.

% \textcolor{red}{Our evaluation mainly demonstrate  We did a partial comparison between Strider and the state of the art RSP engines, \eg C-SPARQL and CQLES.} 
\vspace{-3mm}
\subsection{Evaluation Results \& Discussions}
\label{subsec:eval}
\vspace{-2mm}

\begin{figure}[h]
\vspace{-6.5mm}
\advance\leftskip-0.45cm
\includegraphics[width=1.07\linewidth]{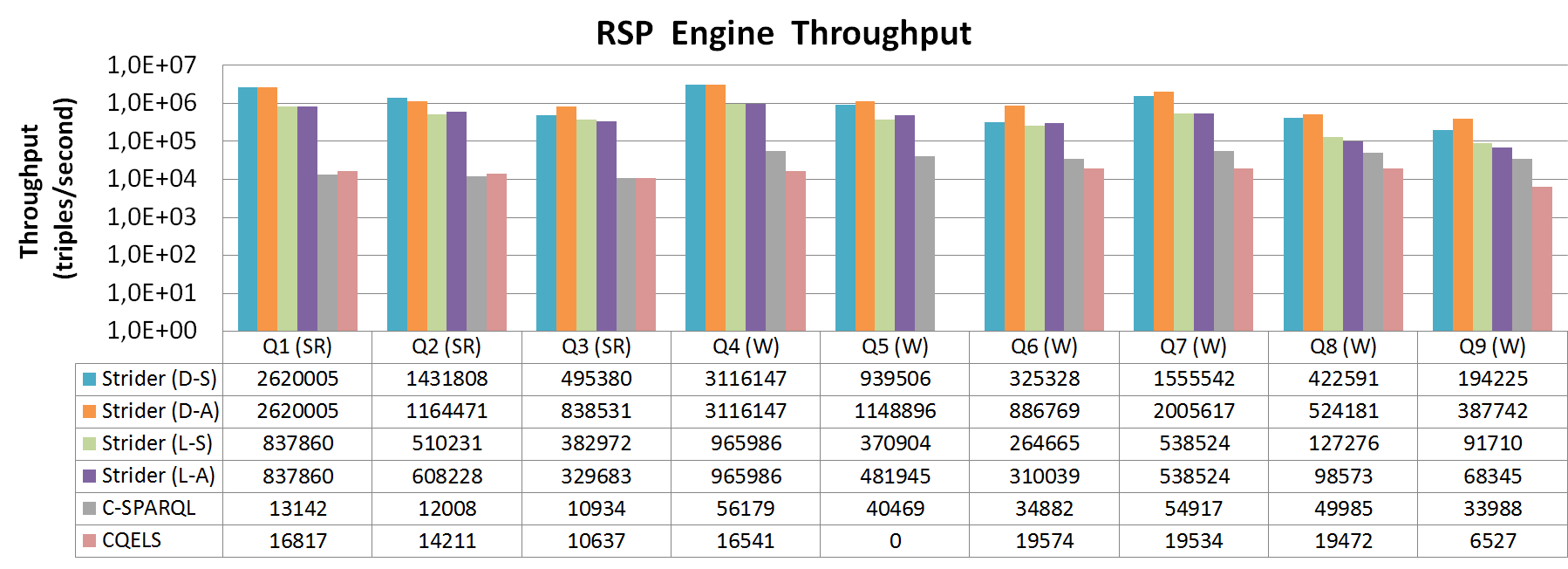}
\captionof{figure}{RSP engine throughput (triples/second). \textbf{D/L-S:} Distributed/Local mode Static Optimization. \textbf{D/L-A:} Distributed/Local mode Adaptive Optimization. \textbf{SR:} Queries for SRBench dataset. \textbf{W:} Queries for Waves dataset.}
\label{fig:throughput}
\vspace{-7.5mm}
\end{figure}

\begin{figure}[h]
\vspace{-5.5mm}
\advance\leftskip-0.55cm
\includegraphics[width=1.07\linewidth]{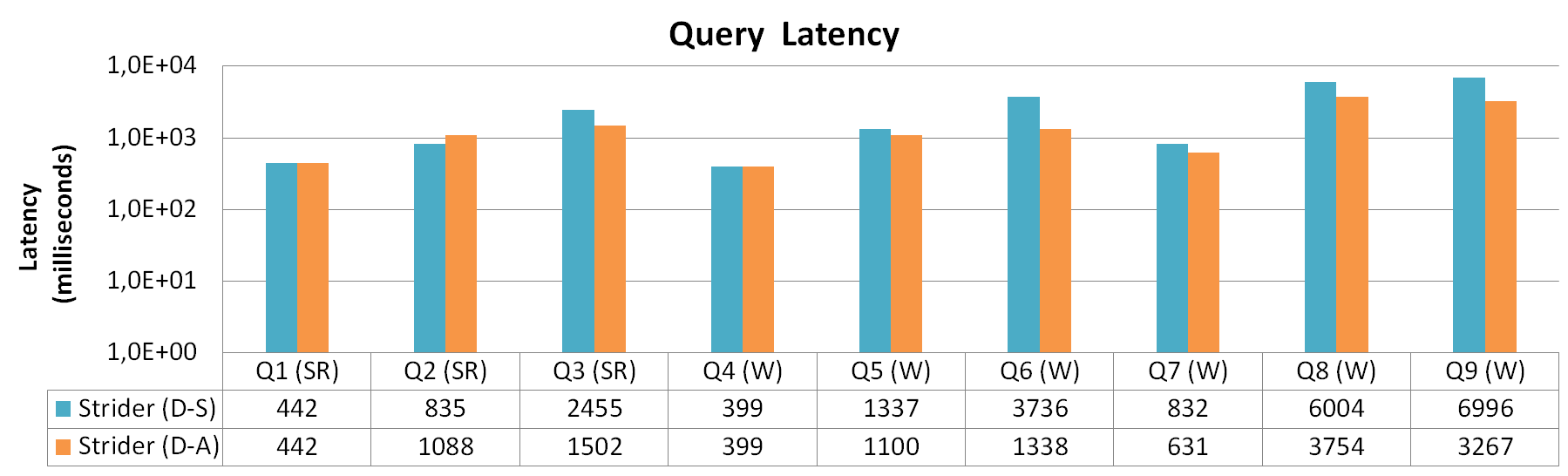}
\captionof{figure}{Query latency (milliseconds) for Strider (in distributed mode)}
\label{fig:latency}
\vspace{-6.5mm}
\end{figure}

Figures \ref{fig:throughput} and \ref{fig:latency} respectively summarize the RSP engines throughput and latency. Note that CQELS gives a parsing error for $Q_5$. This is due, at least for the version that we have evaluated, to the lack of support for the  UNION operator in the source code. In view of the centralized designs of C-SPARQL and CQELS, a direct performance comparison to Strider with distributed hardware settings seems unfair. So we also evaluated Strider in local mode, \ie running the system on a single machine (although it should not be its forte). Based on this preliminary evaluation, we try to give an intuitive impression and reveal our findings about these three RSP systems. 

In Figure \ref{fig:throughput}, we observe that Strider generally achieves million/sub-million-level throughput under our test suite.
Note that both $Q_1$ and $Q_4$ have only one join, \ie optimization is not needed. Most tested queries scale well in Strider. Adaptive optimization generates query plans based on the workload statistics. In total, it provides a more efficient query plan than static optimization. But the gain of AQP for the simple queries that have less join tasks (\eg $Q_1$, $Q_5$) becomes insubstantial. We also found out that, even if Strider runs on a single machine, it still provides up to 60x gain on throughput compared to C-SPARQL and CQELS. Figure \ref{fig:latency} shows Strider attains a second/sub-second delay. Obviously, for queries with 2 triple patterns in the query's BGP, we can observe the same latency between static and adaptive optimizations, $Q_1$ and $Q_4$. Query $Q_2$ is the only query where the latency of the adaptive approach is higher than the static one. This is due to the very simple structure of the BGP (2 joins in the BGP). In this situation, the overhead of DP covers the gain from AQP. For all other queries, the static latency is higher than the adaptive one. This is justified by more complex BGP structures (more than 5 triple patterns per BGP) or some union of BGPs.

On the contrary, the average throughput of C-SPARQL and CQELS is maintained in the range of 6.000 and 50.000 triples/second.  The centralized designs of C-SPARQL and CQELS limit the scalability of the systems. Beyond the implementation of query processing, the reliability of data flow management on C-SPARQL and CQELS could also cause negative impact on system robustness. Due to the lack of some important features for streaming system (\eg back pressure, checkpoint and failure recovery) once input stream rate reaches to certain scale, C-SPARQL and CQELS start behaving abnormally, \eg data loss, exponential increasing latency or query process interruption \cite{LSBench,Mine}. Moreover, we have also observed that CQELS' performance is insensitive to the changing of computing resources. We tested CQELS on different EC2 instance types, \ie with 2, 4 and 8 cores, and the results evaluation variations were negligible. 

\begin{figure}[h]
\vspace{-6.5mm}
\advance\leftskip-0.5cm
\captionsetup{justification=centering,margin=0cm}
\subfloat[\label{subfig-a}]{%
\includegraphics[keepaspectratio=true,scale=0.28]{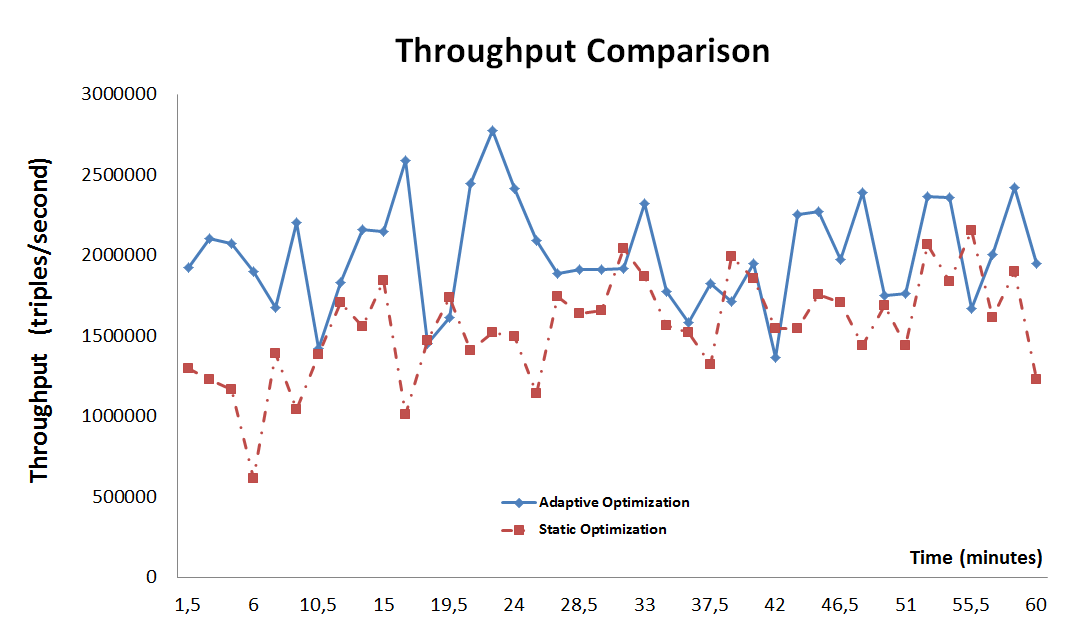}}
\subfloat[\label{subfig-b}]{%
\includegraphics[keepaspectratio=true,scale=0.28]{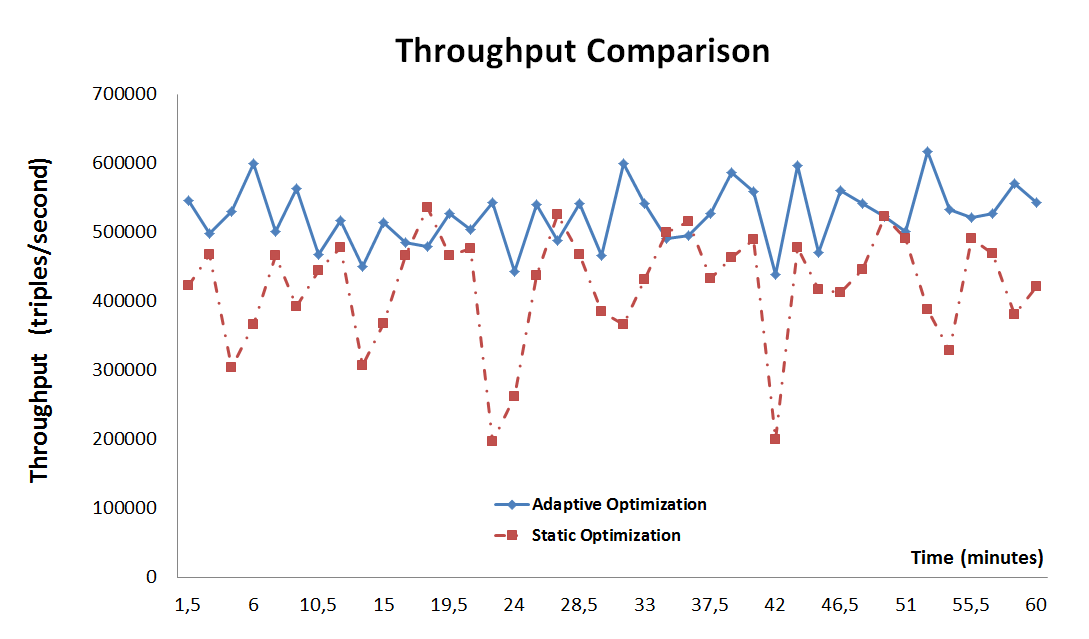}}
\captionof{figure}{Record of throughput on Strider. (a)-throughput for $q_7$; (b)-throughput for $q_8$ }
\label{fig:adpt-5-6}
\vspace{-5.5mm}
\end{figure}

\begin{wrapfigure}{l}{0.52\textwidth}
\vspace{-8.5mm}
  \begin{center}
  \advance\leftskip-0.9cm
    \includegraphics[width=0.6\textwidth]{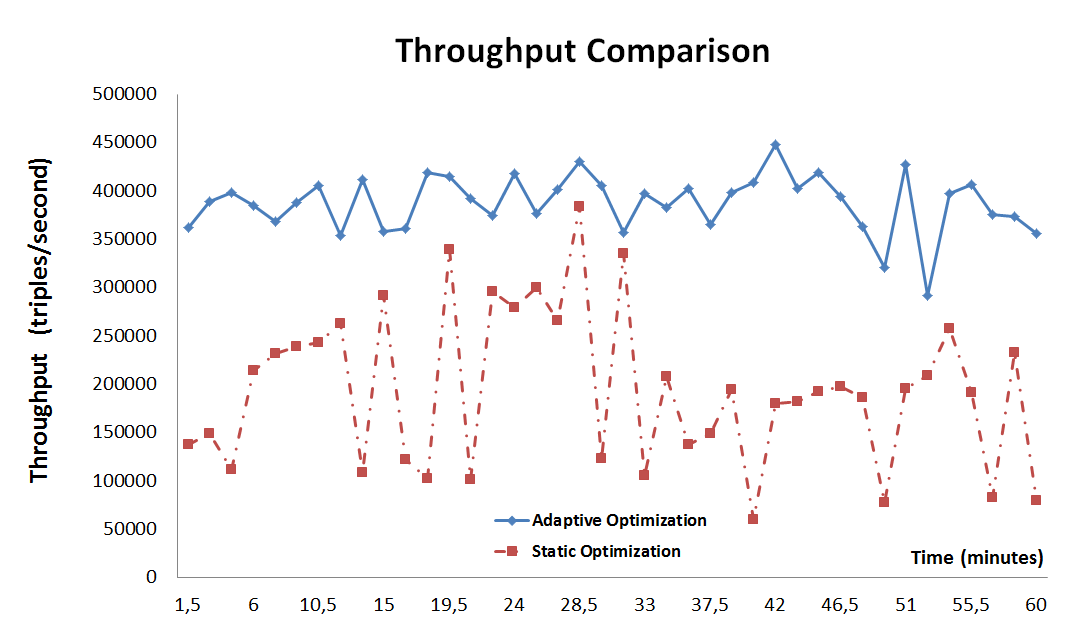}
  \end{center}
  \caption{Throughput for $q_9$ on Strider}
  \label{fig:adpt-7}
  \vspace{-7.5mm}
\end{wrapfigure}

Figure \ref{fig:adpt-5-6} and Figure \ref{fig:adpt-7} concern the monitoring of Strider's throughput for $Q_7$ to $Q_9$. We recorded the changes of throughput over a continuous period of time (one hour). The source stream produces the messages with different types of sensor observations. The stream is generated by mixing temperature, flow and chlorine-level measurement with random proportions. The red and blue curves denote query with respectively  static and adaptive logical plan optimization. For $Q_7$ and $Q_8$ (Figure \ref{fig:adpt-5-6}), except when some serious throughput drops have been observed in \ref{subfig-b}, static and adaptive planners return a close throughput trend. For a more complex query $Q_9$ (Figure \ref{fig:adpt-7}), which contains 9 triple patterns and 8 join operators. Altering logical plans on $Q_9$ causes significant impact on engine performance. Consequently, our adaptive strategy is capable to handle the structurally unstable RDF stream. Thus the engine can avoid a sharp performance degradation.

% \vspace{-7.5mm}

Through this experiment, we identified some shortcomings in Strider that will be addressed in future work: (1) the data preparation on Spark Streaming is relatively expensive. It costs around 0.8 to 1 second to initialize before triggering the query execution in our experiment. (2) Strider has a more substantial throughput decreasing with an increasing number of join tasks. In order to alleviate this effect, the possible solution is enlarging the cluster scale or choosing a more powerful driver node. (3) Strider does not support well high concurrent requests, although this is not at the moment one of our system design goals. \Eg some use cases demand to process a big amount of concurrent queries. Even through Strider allows to perform multiple queries asynchronously, it could be less efficient.

\vspace*{-2mm}
\section{Related Work}
\vspace*{-1mm}
In the recent years, a variety of RSP systems have been proposed which can be divided into two categories: centralized and distributed.

\emph{Centralized RSP engines.} For the last few years, some contributions have been done to satisfy the basic needs of RDF stream processing. RSP engines like C-SPARQL, CQELS, ETALIS, \etc, are developed to run on a single machine. None of them targets the scenario that involves massive incoming data stream.

\emph{Distributed RSP engines.} CQELS-Cloud\cite{CQELSCloud} is the first RSP system which mainly focuses on the engine elasticity and scalability. The whole system is based on Apache Storm. Firstly, CQELS-Cloud compresses the incoming RDF streams by dictionary encoding in order to reduce the data size and the communication in the computing cluster. The query logical plan is mapped to a Storm topology, and the evaluation is done through a series of SPARQL operators located on the vertex of the topology. Then, to overcome the performance bottlenecks on join tasks, the authors propose a \emph{parallel multiway join} based on probing sequence. From the aspect of implementation, CQELS-Cloud is designed as the streaming service for high concurrent requests. The capability of CQELS-Cloud to cope with massive incoming RDF data streams is still missing. Furthermore, to the best of our knowledge, CQELS-Cloud is not open source, customized queries and data feeding are not feasible. Katts is another RSP engine based on Storm.  The implementation of Katts\cite{Katts} is relatively primitive, it is more or less a platform for algorithm testing but not an RSP engine. The main goal of Katts is designed to verify the efficiency of graph partitioning algorithm for cluster communication reduction.

Although the SPARQL query optimization techniques have been well developed recently, CQELS is still the only system which considers query optimization to process RDF data stream. However, the greedy-like left-deep plan leads to sequential query evaluation, which makes CQELS benefit  from few additional computing resources. The conventional SPARQL optimization for static data processing can be hardly applied in a streaming context. Recent efforts \cite{Jena,Thomas-Sigmod,Thomas-CS,S2RDF} possess long data preprocessing stage before launching the query execution. The proposed solutions do not meet real-time or near real-time use cases. The heuristic-based query optimization in \cite{HP} totally ignores data statistics and thus does not promise the optimal execution plan for $24\times7$ running streaming service.

\vspace*{-3mm}
\section{Conclusion and Future Work}
\vspace*{-2mm}
In this paper, we present Strider, a distributed RDF batch stream processing engine for large scale data stream. It is built on top of Spark Streaming and Kafka to support continuous SPARQL query evaluation and thus possesses the characteristics of a production-ready RSP. Strider comes with a set of hybrid AQP strategies: \ie static heuristic rule-based optimization, forward and backward adaptive query processing. We insert the trigger into the optimizer to attain the automatic strategy switching at query runtime. Moreover, with its micro-batch  approach, Strider fills a gap in the current state of RSP ecosystem which solely focuses on record-at-a-time.
%In addition to complement the solution for adaptive SPARQL query processing, \textcolor{green}{explain :} Strider also fills the lack of batch processing gap in the RSP area. 
Through our micro-benchmark based on real-word datasets, Strider provides a million/sub-million-level throughput and second/sub-second latency, a major breakthrough in distributed RSPs. And we also demonstrate the system reliability which is capable to handle the structurally instable RDF streams.

There is still room for improving the system's implementation.
As future work, we aim to add stream reasoning capacities and the ability of combining static data. 
%\textcolor{green}{peut etre pas necessaire: }Furthermore, the data analytics pipeline for anomaly detection is also in the scope of consideration.

\tiny
\bibliographystyle{abbrv}
\bibliography{main}

\begin{thebibliography}{10}

\bibitem{DataFlow}
T.~Akidau, R.~Bradshaw, C.~Chambers, S.~Chernyak, R.~J.
  Fern\'{a}ndez-Moctezuma, R.~Lax, S.~McVeety, D.~Mills, F.~Perry, E.~Schmidt,
  and S.~Whittle.
\newblock The dataflow model: A practical approach to balancing correctness,
  latency, and cost in massive-scale, unbounded, out-of-order data processing.
\newblock {\em PVLDB}, 2015.

\bibitem{City}
M.~I. Ali, F.~Gao, and A.~Mileo.
\newblock Citybench: A configurable benchmark to evaluate rsp engines using
  smart city datasets.
\newblock In {\em ISWC}, 2015.

\bibitem{ETALIS}
D.~Anicic, S.~Rudolph, P.~Fodor, and N.~Stojanovic.
\newblock Stream reasoning and complex event processing in etalis.
\newblock {\em Semant. web}, 2012.

\bibitem{CSPARQL}
D.~F. Barbieri and al.
\newblock {C-SPARQL:} {SPARQL} for continuous querying.
\newblock In {\em WWW}, 2009.

\bibitem{SECRET}
I.~Botan, R.~Derakhshan, N.~Dindar, L.~Haas, R.~J. Miller, and N.~Tatbul.
\newblock Secret: A model for analysis of the execution semantics of stream
  processing systems.
\newblock {\em PVLDB}, 2010.

\bibitem{Flink}
P.~Carbone, A.~Katsifodimos, S.~Ewen, V.~Markl, S.~Haridi, and K.~Tzoumas.
\newblock Apache flink{\texttrademark}: Stream and batch processing in a single
  engine.
\newblock {\em {IEEE} Data Eng. Bull.}, 2015.

\bibitem{Facebook}
G.~J. Chen, J.~L. Wiener, S.~Iyer, A.~Jaiswal, R.~Lei, N.~Simha, W.~Wang,
  K.~Wilfong, T.~Williamson, and S.~Yilmaz.
\newblock Realtime data processing at facebook.
\newblock In {\em SIGMOD}, 2016.

\bibitem{AQP}
A.~Deshpande, Z.~G. Ives, and V.~Raman.
\newblock Adaptive query processing.
\newblock {\em Foundations and Trends in Databases}, 2007.

\bibitem{Katts}
L.~Fischer and al.
\newblock Scalable linked data stream processing via network-aware workload
  scheduling.
\newblock In {\em SSWS@ISWC}, 2013.

\bibitem{Kafka}
K.~Goodhope, J.~Koshy, J.~Kreps, N.~Narkhede, R.~Park, J.~Rao, and V.~Y. Ye.
\newblock Building linkedin’s real-time activity data pipeline.
\newblock {\em IEEE Data Eng. Bull.}, 2012.

\bibitem{Thomas-EDBT}
A.~Gubichev and T.~Neumann.
\newblock Exploiting the query structure for efficient join ordering in sparql
  queries.
\newblock In {\em EDBT}, 2014.

\bibitem{Zookeeper}
P.~Hunt, M.~Konar, F.~P. Junqueira, and B.~Reed.
\newblock Zookeeper: Wait-free coordination for internet-scale systems.
\newblock In {\em USENIX}, 2010.

\bibitem{CQELS}
D.~Le-Phuoc, M.~Dao-Tran, J.~X. Parreira, and M.~Hauswirth.
\newblock A native and adaptive approach for unified processing of linked
  streams and linked data.
\newblock In {\em ISWC}, 2011.

\bibitem{Thomas-CS}
T.~Neumann and G.~Moerkotte.
\newblock Characteristic sets: Accurate cardinality estimation for rdf queries
  with multiple joins.
\newblock In {\em ICDE}, 2011.

\bibitem{Thomas-Sigmod}
T.~Neumann and G.~Weikum.
\newblock Scalable join processing on very large rdf graphs.
\newblock In {\em SIGMOD}, 2009.

\bibitem{Emergent_Schemas}
M.~Pham and P.~A. Boncz.
\newblock Exploiting emergent schemas to make {RDF} systems more efficient.
\newblock In {\em ISWC}, 2016.

\bibitem{CQELSCloud}
D.~L. Phuoc and al.
\newblock Elastic and scalable processing of linked stream data in the cloud.
\newblock In {\em ISWC}, 2013.

\bibitem{LSBench}
D.~L. Phuoc, M.~Dao{-}Tran, M.~Pham, P.~A. Boncz, T.~Eiter, and M.~Fink.
\newblock Linked stream data processing engines: Facts and figures.
\newblock In {\em ISWC}, 2012.

\bibitem{Mine}
X.~Ren, H.~Khrouf, Z.~Kazi{-}Aoul, Y.~Chabchoub, and O.~Cur{\'{e}}.
\newblock On measuring performances of {C-SPARQL} and {CQELS}.
\newblock In {\em SWIT@ISWC}, 2016.

\bibitem{S2RDF}
A.~Sch\"{a}tzle, M.~Przyjaciel-Zablocki, S.~Skilevic, and G.~Lausen.
\newblock S2rdf: Rdf querying with sparql on spark.
\newblock {\em PVLDB.}, 2016.

\bibitem{sparql2sql}
E.~Siow, T.~Tiropanis, and W.~Hall.
\newblock Sparql-to-sql on internet of things databases and streams.
\newblock In {\em ISWC}, 2016.

\bibitem{Jena}
M.~Stocker, A.~Seaborne, A.~Bernstein, C.~Kiefer, and D.~Reynolds.
\newblock Sparql basic graph pattern optimization using selectivity estimation.
\newblock In {\em WWW}, 2008.

\bibitem{Storm}
A.~Toshniwal, S.~Taneja, A.~Shukla, K.~Ramasamy, J.~M. Patel, S.~Kulkarni,
  J.~Jackson, K.~Gade, M.~Fu, J.~Donham, N.~Bhagat, S.~Mittal, and D.~Ryaboy.
\newblock Storm@twitter.
\newblock SIGMOD, 2014.

\bibitem{HP}
P.~Tsialiamanis, L.~Sidirourgos, I.~Fundulaki, V.~Christophides, and P.~Boncz.
\newblock Heuristics-based query optimisation for sparql.
\newblock In {\em EDBT}, 2012.

\bibitem{Drizzle}
S.~Venkataraman, A.~Panda, K.~Ousterhout, A.~Ghodsi, M.~J. Franklin, B.~Recht,
  and I.~Stoica.
\newblock Drizzle: Fast and adaptable stream processing at scale.
\newblock In {\em Spark Summit}, 2016.

\bibitem{Spark}
M.~Zaharia, M.~Chowdhury, T.~Das, A.~Dave, J.~Ma, M.~McCauley, M.~J. Franklin,
  S.~Shenker, and I.~Stoica.
\newblock Resilient distributed datasets: A fault-tolerant abstraction for
  in-memory cluster computing.
\newblock In {\em NSDI}, 2012.

\bibitem{Sparkstreaming}
M.~Zaharia, T.~Das, H.~Li, T.~Hunter, S.~Shenker, and I.~Stoica.
\newblock Discretized streams: Fault-tolerant streaming computation at scale.
\newblock In {\em SOSP}, 2013.

\bibitem{SRBench}
Y.~Zhang, P.~M. Duc, O.~Corcho, and J.-P. Calbimonte.
\newblock Srbench: A streaming rdf/sparql benchmark.
\newblock In {\em ISWC}, 2012.

\end{thebibliography}
\end{document}